\documentclass[11pt]{article}
\usepackage{amsmath, amsthm}
\usepackage{newpxtext}
\usepackage{newpxmath}
\usepackage{tikz}
\usepackage{float}
\usepackage{stackrel}
\usepackage{hyperref}
\usepackage{mathtools}
\usepackage{comment}
\usepackage[margin=1in]{geometry}

\usepackage{algorithm}
\usepackage{algpseudocode}
\usepackage{accents}

\usepackage{bm}

\usepackage{fancybox}
\usepackage{color}
\usepackage{latexsym}

\usepackage{eepic}
\usepackage{epic}
\usepackage{epsf}
\usepackage{graphics}
\usepackage{dsfont}

\newtheorem{theorem}{Theorem}
\newtheorem*{theorem*}{Theorem}

\newtheorem{lemma}[theorem]{Lemma}

\newtheorem{fact}[theorem]{Fact}
\newtheorem{corollary}[theorem]{Corollary}
\newtheorem{claim}[theorem]{Claim}

\newtheorem{proposition}[theorem]{Proposition}
\theoremstyle{definition}

\newtheorem{definition}[theorem]{Definition}
\newtheorem{remark}[theorem]{Remark}
\newenvironment{claimproof}[1]{\par\noindent{\em Proof of Claim.}~#1}

\newcommand{\B}{\mathfrak{B}}

\newcommand{\M}{\mathbf{M}}
\renewcommand{\r}{\mathfrak{r}}

\newcommand{\F}{\mathds{F}}

\newcommand{\Q}{\mathds{Q}}

\DeclareMathOperator{\ncrank}{\mathrm ncrank}

\DeclareMathOperator{\poly}{poly}
\DeclareMathOperator{\rank}{rank}

\DeclareMathOperator{\Gal}{Gal}

\DeclareMathOperator{\sing}{\rm SINGULAR}
\DeclareMathOperator{\nsing}{\rm NSINGULAR}

\DeclareMathOperator{\diag}{diag}

\DeclareMathOperator{\cir}{circ}
 
\newcommand{\p}{\mbox{\small\rm P}}

\renewcommand{\ge}{\geqslant}
\renewcommand{\geq}{\geqslant}
\renewcommand{\le}{\leqslant}
\renewcommand{\leq}{\leqslant}

\DeclareFontFamily{OMX}{MnSymbolE}{}
\DeclareFontShape{OMX}{MnSymbolE}{m}{n}{
   <-6>  MnSymbolE5
   <6-7>  MnSymbolE6
   <7-8>  MnSymbolE7
   <8-9>  MnSymbolE8
   <9-10> MnSymbolE9
   <10-12> MnSymbolE10
   <12->   MnSymbolE12}{}
\DeclareSymbolFont{mnlargesymbols}{OMX}{MnSymbolE}{m}{n}
\SetSymbolFont{mnlargesymbols}{bold}{OMX}{MnSymbolE}{b}{n}
\DeclareMathDelimiter{\llangle}{\mathopen}{mnlargesymbols}{'164}{mnlargesymbols}{'164}
\DeclareMathDelimiter{\rrangle}{\mathclose}{mnlargesymbols}{'171}{mnlargesymbols}{'171}

\renewcommand{\angle}[1]{\langle #1 \rangle}
\newcommand{\ubar}[1]{\underaccent{\bar} #1}

\DeclareFontEncoding{LS1}{}{}
\DeclareFontSubstitution{LS1}{stix}{m}{n}
\DeclareSymbolFont{symbols2stix}{LS1}{stixfrak} {m} {n}
\DeclareMathSymbol{\lparenless}{\mathopen} {symbols2stix}{"32}
\DeclareMathSymbol{\rparengtr}{\mathclose}{symbols2stix}{"33}

\newcommand{\newbrak}[1]{{\lparenless} #1 {\rparengtr}}

\title{The Noncommutative Edmonds' Problem Re-visited}
\author{
Abhranil Chatterjee\thanks{Indian Statistical Institute, Kolkata, \texttt{email: abhneil@gmail.com}} 
\and Partha Mukhopadhyay\thanks{Chennai Mathematical Institute, Chennai, \texttt{email: partham@cmi.ac.in}}
}

\date{}
\begin{document}
\maketitle

\begin{abstract}
 Let $T$ be a matrix whose entries are linear forms over the noncommutative variables $x_1, x_2, \ldots, x_n$. The noncommutative Edmonds' problem ($\nsing$) aims to determine whether $T$ is invertible in the free skew field generated by $x_1,x_2,\ldots,x_n$. Currently, there are three different deterministic polynomial-time algorithms to solve this problem: using operator scaling~\cite{GGOW16}, algebraic methods~\cite{IQS18}, and convex optimization~\cite{HH21}. 
 
 In this paper, we present a simpler algorithm for the $\nsing$ problem. While our algorithmic template is similar to the one in \cite{IQS18}, it significantly differs in its implementation of the \emph{rank increment} step. Instead of computing the limit of a second Wong sequence, we reduce the problem to the polynomial identity testing (PIT) of noncommutative algebraic branching programs (ABPs).
 
 This enables us to bound the bit-complexity of the algorithm over $\Q$ without requiring special care. 
 Moreover, the rank increment step can be implemented in quasipolynomial-time even without an explicit description of the coefficient matrices in $T$. This is possible by exploiting the connection with the black-box PIT of noncommutative ABPs \cite{FS13}. % Moreover, our algorithm does not need an explicit description of the coefficient matrices in performing the rank increment step, exploiting the connection with the black-box PIT of noncommutative ABPs. 
 %It could be useful for understanding the black-box and the parallel complexity of $\nsing$, which are long-standing open problems. 
\end{abstract}

\newpage
\section{Introduction}
Let $\ubar{x}=\{x_1, \ldots, x_n\}$ be $n$ variables and $\F$ be a field. Consider the coefficient matrices $A_0, A_1, \ldots, A_n \in \M_s(\F)$, and define an $s \times s$ symbolic matrix $T$ as $T = A_0 + A_1x_1+\ldots + A_nx_n$.
In 1967, Edmonds introduced the problem of deciding whether $T$ is invertible over the rational function field $\F(x_1, x_2, \ldots, x_n)$~\cite{Edm67}, often referred to as the $\sing$ problem.
More generally, Edmonds was interested in computing the (commutative) rank of $T$ over the rational function field $\F(x_1, \ldots, x_n)$. Equivalently, the problem asks to compute the maximum rank of a matrix in the matrix space generated by the $\F$-linear span of the coefficient matrices. The problem was further studied by Lov\'{a}sz~\cite{Lov89}, in the context of graph matching and matroid-related problems. The $\sing$ problem, more generally the rank computation problem, admits a simple randomized polynomial-time algorithm due to the %Schwartz-Zippel 
Polynomial Identity Lemma~\cite{Sch80, Zip79, DL78}. However, the quest for an efficient \emph{deterministic} algorithm remains elusive. Eventually, Kabanets and Impagliazzo showed that any efficient deterministic algorithm for $\sing$ would lead to a strong circuit lower bound, justifying the elusiveness over the years \cite{KI04}. Interestingly, the commutative rank computation problem admits a deterministic PTAS algorithm~\cite{BJP18}.

%\textcolor{red}{In recent times, several groups of researchers have studied the complexity of this problem in the noncommutative setting.}
The rank computation problem is also well-studied in the noncommutative setting~\cite{Coh95, FR04}.
More precisely, $T$ is still a linear matrix but the variables $x_1, \ldots, x_n$ are noncommuting. The problem of testing whether $T$ is invertible ($\nsing$), or the rank computation question is naturally addressed over the noncommutative analog of the commutative function field, \emph{the free skew field} $\F\newbrak{\ubar{x}}=\F\newbrak{x_1, x_2, \ldots, x_n}$. %It takes some work to formally define the free skew field
The free skew field has been extensively studied in mathematics~\cite{Ami66, ami55, Coh71} and the definition is somewhat involved.
%complicated. 
However, for the purpose of this paper, it is enough to state that
$\F\newbrak{\ubar{x}}$ is the smallest field over the variables $\{x_1, x_2, \ldots, x_n\}$ when we drop even the commutativity relations. %This is the noncommutative analog of the commutative function field $\F(\ubar{x})$. 
Similar to the commutative setting, the noncommutative rank computation of $T$ has an equivalent definition involving the matrix space generated by the $\F$-linear span of the coefficient matrices. We say that the noncommutative rank of $T$ is $s-c$ if $c$ is the maximum integer such that there exists a \emph{$c$-shrunk subspace}\footnote{We say $U\leq \F^n$ is a $c$-shrunk subspace of a matrix space $\mathcal{B}$ if there exists $W\ \leq \F^n$ such that $\dim(W) \leq \dim(U) - c$ and for all $B\in \mathcal{B}$, the set $\{Bu:u\in U\}\leq W.$} of the said matrix space.  %via matrix blow-up spaces, which we discuss in Section \ref{sec:overview}. 

Two independent breakthrough results showed that $\nsing\in \p$ \cite{GGOW16, IQS18}. In particular, the algorithm of Garg, Gurvits, Oliveira, and Wigderson \cite{GGOW16} is analytic in nature and based on operator scaling which works over $\Q$. The algorithm of Ivanyos, Qiao, and Subrahmanyam \cite{IQS18} is purely algebraic. 
%It is based on ideas stimulated by \textcolor{red}{invariant theory} and the regularity lemma \cite{DM17}. 
Moreover, the algorithm in their paper~\cite{IQS18} works over $\Q$ and fields with positive characteristics. Subsequently, a third algorithm based on convex optimization is also developed by Hamada and Hirai \cite{HH21}. Not only these are beautiful results, but also they enriched the field of computational invariant theory greatly \cite{BFGOWW19, DM20}.  
%The noncommutative version of Edmonds’ problem asks to find the noncommutative rank of $T$ over the free skew field $\F\newbrak{x_1, \ldots, x_n}$.
%In the commutative setting, this problem is solvable in randomized polynomial time due to the Polynomial Identity Lemma. Derandomizing Edmonds' problem is a challenging open problem in theoretical computer science which also captures the PIT of ABPs. However, it admits a deterministic PTAS~\cite{BJP18}. Interestingly, the noncommutative counterpart of the problem admits a deterministic polynomial-time algorithm~\cite{GGOW16, IQS18}.

In this paper, we propose a simpler algorithm for $\nsing$. The main algorithmic template of our algorithm is similar to \cite{IQS18}. However, there is an important difference in implementing one of the core steps that we explain in the next subsection.  
\begin{theorem}\label{thm:main-theorem}
Given an $s\times s$ matrix $T$ whose entries are $\Q$-linear forms over the noncommuting variables $\{x_1, \ldots, x_n\}$, the noncommutative rank of $T$ over $\Q\newbrak{x_1, \ldots, x_n}$ can be computed in deterministic $\poly(s,n)$ time. As a special case, $\nsing\in\p$.
\end{theorem}

\begin{remark}\label{rmk:allfields}
The result of \cite{IQS18} works over fields of positive characteristics using the rudiments of Galois theory. Similarly, our algorithm can also be extended over fields of positive characteristics. Since the key algorithmic ideas remain exactly the same, we prefer to describe the algorithm only over $\Q$ to minimize the use of the Galois theory machinery.  
\end{remark}

\subsection{The Overview}\label{sec:overview}
The noncommutative rank ($\ncrank$) of an $s\times s$ matrix $T$ over the free skew field is the minimum $r$ such that $T$ can be written as $T=PQ$ where $P$ and $Q$ are $s\times r$ and $r\times s$ matrices with linear entries. This is also referred to as the inner rank of $T$~\cite{Coh95}. There are several equivalent definitions of noncommutative rank (see \cite{GGOW20, IQS18} for more details). The definition of our particular interest in this paper is the blow-up definition \cite{IQS18, DM17}. Let $T$ be written as $T = A_0 + \sum_{i=1}^n A_i x_i$ where $A_0, A_1, \ldots, A_n$ are the coefficient matrices. For any matrix tuple $\ubar{p}=(p_1, p_2, \ldots, p_n)$ of dimension $d$, the evaluation is defined as the following $sd\times sd$ matrix: 
\[
T(\ubar{p})= A_0\otimes I_d + \sum_{i=1}^n A_i\otimes p_i.
\]
Define $T^{\{d\}}$ as the set of $sd\times sd$ matrices such that for each $B\in T^{\{d\}}$, there is a $d$-dimensional matrix tuple $\ubar{p}$ such that $B = T(\ubar{p})$. Let $\rank(T^{\{d\}})$ be the maximum rank attained by a matrix in $T^{\{d\}}$. The regularity lemma \cite{IQS18, DM17} shows that $\rank(T^{\{d\}})$ is a multiple of $d$. The blow-up definition of the noncommutative rank 
is the limit of the sequence $\left(\frac{\rank(T(\ubar{p}))}{d}\right)$ as $d\rightarrow \infty$. It follows from \cite{FR04, IQS18} that
\[\ncrank(T) = \underset{d\rightarrow \infty}{\rm lim}\left(\frac{\rank(T(\ubar{p}))}{d}\right).\] 

In this paper, we present a deterministic polynomial-time algorithm to compute the noncommutative rank of a linear matrix. The algorithmic template is similar to the algorithm of Ivanyos, Qiao, and Subrahmanyam~\cite{IQS18}. Before we present our algorithm, we first go over the key steps of the algorithm presented in~\cite{IQS18}. Given a linear matrix $T$ of size $s$, their algorithm gradually computes the rank of $T$. 
%Their algorithm is an incremental one where given a linear matrix $T$, 
Suppose the algorithm outputs a matrix in $T^{\{d\}}$ of rank at least $rd$ at any intermediate stage (where it is always maintained that $d\leq s+1$). The next stage consists of two main steps: (a) rank increment step, (b) rounding and blow-up control.

%More precisely, the algorithm outputs a matrix in $T^{\{d\}}$ of rank at least $rd$ (where it is always maintained that $d\leq r+1$)  at the $r^{th}$ stage. Each stage consists of two main steps: (a) rank increment step, (b) rounding and blow-up control.

\begin{description}
\item[(a) Rank Increment Step]
\end{description}
Given a matrix $B$ in $T^{\{d\}}$ of rank $\geq rd$,
%(by construction, $d \leq r+1$), 
the algorithm first checks whether there exists another matrix $B'$ in $T^{\{d'\}}$ (where $d' > d$) of rank $\geq rd' + 1$. If no such matrix exists, we output $\ncrank(T) = r$. The correctness follows from the blow-up definition. Otherwise, the algorithm finds such a $B'$ and proceeds to the next step. This step is technically involved and uses a linear algebraic procedure involving the limit point computation of a \emph{second Wong sequence} \cite{IKQS15, IQS18}. Informally speaking, this is analogous to the augmenting-path algorithm for bipartite matching.  
%analogue of the augmenting paths, the \emph{second Wong sequence} \cite{IKQS15, IQS18}. 
A direct implementation of the limit point computation incurs a bit-complexity blow-up over $\Q$ due to the repeated application of Gaussian eliminations. To tackle this, the notion of matrix pseudo-inverse is used \cite{IKQS15, IQS18}.  
%Moreover, it requires extra care over the rationals to tackle the bit-complexity blow-up over $\Q$.

%it involves a bit-complexity blow up over $\Q$ due to repeated application of Gaussian elimination. To tackle this, the subtle concept of the matrix pseudo-inverse is used \cite{IKQS15, IQS18}.

%The key step in \cite{IQS18} is a procedure using the second  Wong sequence to produce another 
%witness from the \emph{image} of $T$ such that the rank is $>rd$. 
%tuple $\ubar{q}$ (of possibly higher dimension) such that $\rank(T(\ubar{q}))>rd$. 
%This is called the \emph{rank-increment step}. Very intuitively, the application of the second Wong sequence is an operator theoretic generalization of the augmenting-path algorithm for bipartite matching \cite{IKQS15, IQS18}. Computing the second Wong sequence is a highly sequential procedure and also it involves bit-complexity blow up (over $\Q$) due to repeated application of Gaussian elimination. To tackle this, the subtle concept of the matrix pseudo-inverse is used \cite{IKQS15, IQS18}. 

\begin{description}
\item[(b) Rounding and Blow-up Control]
\end{description}
Once we have the matrix $B'$ in $T^{\{d'\}}$ of rank $\geq rd' + 1$, the rounding step of the algorithm uses a constructive version of the regularity lemma to find another matrix $B''$ in $T^{\{d'\}}$ such that the rank of $B''$ is a multiple of $d'$ and $\rank(B'') \geq \rank(B')$. That implies the rank of $B''$ is $r'd'$ where $r'$ is at least $r + 1$. 
Here, $d'$ is at least a constant multiple of $d$. One can not afford such a blow-up in the dimension during every round of the algorithm as it incurs an exponential blow-up in the final dimension. They reduce the dimension by repeated application of the rounding step and the constructive version of the regularity lemma. Finally, it outputs a matrix $\hat{B}$ of rank $r'd''$ where $r' \geq r + 1$ and $d''\leq r'+1$.
%Finally, it outputs a matrix $\hat{B}$ in $T^{\{r'+1\}}$ of rank $\geq r'(r'+1)$ by repeated application of the rounding step and the constructive version of the regularity lemma. This is important because the parameter $d'$ is at least a constant multiple of $d$ and one can not afford such blow-up in the dimension during every round of the algorithm. 

\begin{description}
\item [Our algorithm] 
\end{description}
 One of the main features of our algorithm is that, we implement the rank increment step without using the second Wong sequence. At a high level, the rank increment step of our algorithm has a conceptual similarity with the proof idea used in~\cite{BBJP19}. If for a matrix tuple $\ubar{p}$ of dimension $d$, the rank of $T(\ubar{p})\geq rd$, then we say that $\ubar{p}$ is a witness of noncommutative rank $r$ of $T$. Using simple linear algebraic ideas, we reduce the rank increment step to the polynomial identity testing of noncommutative ABPs. As we show in Lemma \ref{lem:trucated}, Lemma \ref{lem:nonzero-output}, and Corollary \ref{cor:immediate} that to find a witness of a larger rank, it suffices to compute a matrix tuple where a noncommutative ABP does not evaluate to zero. This requires applying simple and well-known identity testing algorithms \cite{RS05, AMS10}. This way one can compute a matrix tuple $\ubar{p'}$ of dimension $d_1$ such that the rank of $T(\ubar{p'}) > rd_1$. Interestingly, there is no intermediate bit-complexity blow-up over $\Q$. 

The rest of the algorithm follows the rounding and the blow-up control steps of \cite{IQS18} closely. In Lemma \ref{lem:const-reg}, we show how to embed $\ubar{p'}$ in a division algebra $D$ of index $d_1$ to compute another tuple $\ubar{p''}$, a witness of rank $r'$ of $T$ where $r' \geq r+1$. More precisely, the rank of $T(\ubar{p''}) \geq (r+1)d_1$. For the division algebra $D$, we use a well-known explicit construction of cyclic division algebra from~\cite{Lam01}. 

We then use Lemma \ref{lem:blow-up-control} to control the blow-up in the dimension and output the matrix tuple $\widehat{\ubar{p}}$ of dimension at most $d'$ such that $\rank(T)\geq (r+1)d'$ where $d'\leq r'+1$. 
Hence $\widehat{\ubar{p}}$ is a witness that $\ncrank(T)\geq r+1$. 
We need to repeat the entire procedure for at most $s$ rounds where $s$ is the dimension of $T$. 
%The novelty of our algorithm is in the implementation of the rank increment step where we avoid the computation of the limit point of a second Wong sequence. Given a matrix $B$ in $T^{\{d\}}$ of rank $\geq rd$, we reduce the problem of deciding whether there exists another matrix $B'$ in $T^{\{d'\}}$ (where $d' > d$) of rank $\geq rd' + 1$ to polynomially many instances of polynomial identity testing for noncommutative algebraic branching programs (ABPs) that admits a simple deterministic polynomial-time algorithm~\cite{RS05, AMS10}. 
%This is the step where our algorithm works very differently. It avoids computing the second Wong sequence and reduce the rank increment step to polynomial identity testing for noncommutative algebraic branching programs (ABPs) \cite{RS05, AMS10} which has efficient (and simple) deterministic polynomial-time algorithm. 
%In particular, the algorithm computes a $d'\times d'$ matrix tuple $\ubar{q} = (q_1, \ldots, q_n)$ such that $\rank(T(\ubar{q}))\geq rd' + 1$. 
%Interestingly, there is no bit-complexity blow-up in our algorithm. 

We finally remark that 
%Another interesting feature of our algorithm is that 
unlike \cite{IQS18}, the rank increment step does not need an explicit description of the coefficient matrices of $T$ if we settle with a quasipolynomial-time algorithm. Essentially, we can use the quasipolynomial-size explicit hitting set construction for the identity testing of noncommutative ABPs by Forbes and Shpilka \cite{FS13} in place of the algorithm by Raz and Shplika \cite{RS05}. We elaborate on this 
%As we discuss 
in Section \ref{sec:conclusion}.
We believe that the connection with the noncommutative polynomial identity testing  could be useful 
%this gives us some hope of  
in understanding the black-box and the parallel complexity of $\nsing$ which are long-standing open problems \cite{GGOW16, GGOW20}.\\

%\subsection*{Organization} 
\noindent\textbf{Organization.} In Section \ref{sec:prelim}, we collect background results from algebraic complexity theory and cyclic division algebras. We prove Theorem \ref{thm:main-theorem} in Section \ref{sect-idea}. The main contribution of the paper is in the implementation of the rank increment step which is presented in Subsection \ref{sec:rankincrement}. The final algorithm is presented in Subsection \ref{sec:final-algorithm}. Section \ref{sec:conclusion} contains some additional remarks related to the rank increment step in the black-box setting.  
%The rounding and blow-up control step of our algorithm is very similar to that of \cite{IQS18} where we use a well-known explicit construction of cyclic division algebra from~\cite{Lam01}. %Finally, there is a simple step to ensure that the blow-up dimension $d$ can always be reduced within $s$ throughout the algorithm. 
\section{Background and Notation}\label{sec:prelim} 

%\subsection*{Basic notation.}
Throughout the paper, we use $\F, F, K$ for fields. ${\M_m(\F)}$ (resp. ${\M_m(F)}, {\M_m(K)}$) is used for $m$ dimensional matrix algebra over $\F$ where $m$ is clear from the context. Similarly, we use ${\M_m(\F)}^n$ (resp. ${\M_m(F)}^n, {\M_m(K)}^n$) to denote the set of $n$ tuples over ${\M_m(\F)}$ (resp. ${\M_m(F)}, {\M_m(K)}$).  
%of $m$ dimensional matrix algebra over $\F$. 
$D$ is used to denote a division algebra. %and $D$ be a cyclic division algebra of certain index. 
Let $\ubar{x}$ be the set of variables $\{x_1, \ldots, x_n\}$. 
Sometimes we use $\ubar{p}, \ubar{q}$ to denote the matrix tuples in suitable matrix algebras. The free noncommutative ring of polynomials over a field $\F$ is denoted by $\F\angle{\ubar{x}}$. 
%The ring of \emph{formal power series} is denoted by $\F\dangle{\ubar{x}}$. 
The notation $A\otimes B$ is the usual tensor product of the matrices $A,B$.   
%For a series (or polynomial) $S$, the coefficient of a monomial (word) in $S$ is denoted by $[m]S$. The formal power series over $\M_m(\F)$ is denoted by $\M_m(\F)\dangle{\ubar{x}}$.  
%We use $\Supp(S)$ to denote the \emph{support} of the series $S$: $ \Supp(S)=\{m\mid [m]S\ne 0\}$.
%Noncommutative monomials are also words. 

%$\F\angle{\ubar{x}}$ is the $\F$-algebra of noncommutative polynomials and $R_{\F}\newbrak{x}$ is the set of rational expressions. For a rational expression $r$, let $\dom(r)$ be the set of matrix tuple (of any dimension) where $r$ is defined. 

\subsection{Algebraic Complexity} 

%We first recall the definition of an algebraic branching program.

\begin{definition}[Algebraic Branching Program]\label{abpdefn}
An \emph{algebraic branching program} (ABP) is a layered directed
acyclic graph. The vertex set is partitioned into layers
$0,1,\ldots,d$, with directed edges only between adjacent layers ($i$
to $i+1$). There is a \emph{source} vertex of in-degree $0$ in the layer
$0$, and one out-degree $0$ \emph{sink} vertex in layer $d$. Each edge
is labeled by an affine $\F$-linear form. The polynomial computed by
the ABP is the sum over all source-to-sink directed paths of the
ordered product of affine forms labeling the path edges. 
\end{definition}

The \emph{size} of the ABP is defined as the total number of nodes and the \emph{width} is the maximum number of nodes in a
layer. An ABP can compute a commutative or a noncommutative polynomial. ABPs of width $w$ can also be seen as iterated matrix multiplication $ \ubar{c}\cdot M_1 M_2 \cdots M_{\ell} \cdot\ubar{b} $, where $\ubar{c}, \ubar{b}$ are $1\times w$ and $w \times 1$ vectors respectively and each $M_i$ is a $w \times w$ matrix, whose entries are affine linear forms over $\ubar{x}$.

%We also consider commutative set-multilinear ABPs and read-once oblivious ABPs (ROABPs). For the set-multilinear case, the
%(commutative) variable set is partitioned as $Y = Y_1\sqcup
%Y_2\sqcup\cdots \sqcup Y_d$ where for each $j\in [d]$, $Y_j =
%\{y_{ij}\}_{i=1}^n$. An ABP $B$ is a homogeneous set-multilinear if each edge in the $j^{th}$ layer of the ABP is labelled by linear forms over $Y_j$. For ROABP, a different variable is used for each layer, and the edge labels are univariate polynomials. Therefore, an ROABP of $d$ layers can be represented as $ \boldsymbol{c} \cdot M_1(v_1) M_2(v_2) \cdots M_{v_d}(d) \cdot\boldsymbol{b}$. We say that the ROABP respects the variable order $v_1 < v_2 < \cdots < v_d$.

%Given a homogeneous degree $d$ noncommutative polynomial $f$, its set-multilinearization $\sm(f)$ is the corresponding set-multilinear polynomial obtained by replacing $x_i$ in the $j^{th}$ position (in a monomial) by $y_{i,j}$ in every monomial. Clearly, $f\equiv 0$ if and only if $\sm(f)\equiv 0$.

%\vspace{1mm}
%\subsubsection*{Identity testing results}
\begin{description}
\item[Identity testing results] 
\end{description}
Given a noncommutative ABP, Raz and Shpilka have given a deterministic polynomial time algorithm to check whether the polynomial computed by the ABP is zero or not \cite{RS05}. 

\begin{theorem}[Raz-Shpilka \cite{RS05}]\label{razshpilka}
Given a noncommutative ABP of width $w$ and $d$ many layers computing a polynomial $f \in \F \angle{\ubar{x}}$, there is a deterministic $\poly(w,d,n)$ time algorithm to test whether $f \equiv 0$ or not. 
\end{theorem} 
In fact, the following corollary is standard by now. This was first formally observed in \cite{AMS10} using a minor adaptation of \cite{RS05}. 

\begin{corollary}\label{cor:rs-ams-adapted}
Given a noncommutative ABP of width $w$ and $d$ many layers computing a nonzero polynomial $f \in \F \angle{\ubar{x}}$, there is a deterministic $\poly(w,d,n)$ time algorithm which outputs a nonzero monomial $m$ in $f$. If $\F=\Q$, the bit-complexity of the algorithm is $\poly(w,d,n,b)$ where $b$ is the maximum bit-complexity of any coefficient in the input ABP.  
\end{corollary}

Essentially, the algorithm of Raz and Shpilka maintains basis vectors (indexed by at most $w$ monomials) in each layer of the ABP using simple linear algebraic computations. The entries of the basis vectors are the coefficients of the indexing monomials in different nodes of the ABP along the width. 

Given such a monomial $m=x_{i_1} x_{i_2} \ldots x_{i_d}$, \cite{AMS10} introduced a simple trick to produce a matrix tuple in 
$\M_{d+1}(\F)^n$ on which $f$ evaluates to nonzero. To see that consider a $d+1$ state deterministic finite automaton $\mathcal{A}$ that accepts only the string $x_{i_1} x_{i_2} \ldots x_{i_d}$ over the alphabet $\{x_1, x_2, \ldots, x_n\}$. The transition matrix tuple 
$(M_{x_1}, \ldots, M_{x_n})$ of $\mathcal{A}$ have the property that $f(M_{x_1}, \ldots, M_{x_n})\neq 0$. More precisely, the automaton $\mathcal{A}$ is the following.

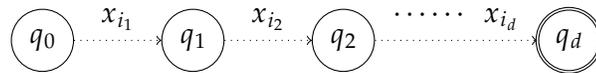
\begin{figure}[H]
\begin{center}
\begin{tikzpicture}
\node(pseudo) at (-1,0){};
\node(0) at (0,0)[shape=circle,draw]        {$q_0$};
\node(1) at (2,0)[shape=circle,draw]        {$q_1$};
\node(2) at (4,0)[shape=circle,draw]        {$q_2$};
\node(3) at (7,0)[shape=circle,draw,double] {$q_{d}$};
\path [->]
  (0)      edge                 node [above]  {$x_{i_1}$}     (1)
  (1)      edge                 node [above]  {$x_{i_2}$}     (2)
  (2)   [dotted]   edge                 node [above]  {$\cdots \cdots ~~x_{i_d}$}     (3);
  %(2)      edge [bend left=30]  node [below]  {a}     (0)
  %(0)      edge [loop above]    node [above]  {$\xi_1$}     ()
  %(1)      edge [loop above]    node [above]  {$\xi_2$}     ()
%(2)      edge [loop above]    node [above]  {$\xi_3$}     ()
 % (3)      edge [loop above]    node [above]  {$\xi_{k +1}$}   ()
  %(pseudo) edge                                       (0);
\end{tikzpicture}
\end{center}
\end{figure}

The transition matrices $M_{x_j} : 1\leq j\leq n$ are $(d+1)$ dimensional $(0,1)$-matrices with the property that $M_{x_j}(\ell,\ell+1)=1$ if and only if $x_j$ is the edge label between $q_{\ell}$ and $q_{\ell+1}$ for $0\leq \ell\leq d-1$.  
This we record as a corollary. 

\begin{corollary}\label{cor:abp-witness}
Given a noncommutative ABP of width $w$ and $d$ layers computing a nonzero polynomial $f\in\F\angle{\ubar{x}}$, there is a deterministic polynomial-time algorithm that can output a matrix tuple $(M_1, M_2, \ldots, M_n)$ of dimension at most $d+1$ such that $f(M_1, M_2,\ldots, M_n)\neq 0$. 
\end{corollary}

\subsection{Cyclic Division Algebras}\label{sec:cyclic}
A division algebra $D$ is an associative algebra over a (commutative) field $\F$ such that all 
nonzero elements in $D$ are units (they have a multiplicative inverse). In the context of this 
paper, we are interested in finite-dimensional division algebras. Specifically, we focus on cyclic division algebras and their construction \cite[Chapter 5]{Lam01}. Let $F=\Q(z)$, where $z$ is a commuting indeterminate. Let $\omega$ be an $\ell^{th}$ primitive root of unity. To
be specific, let $\omega= e^{2\pi i/\ell}$. Let
$K=F(\omega)=\Q(\omega,z)$ be the cyclic Galois extension of $F$ obtained by
adjoining $\omega$. The elements of $K$ are polynomials in $\omega$ (of
degree at most $\ell-1$) with coefficients from $F$.

Define $\sigma:K\to K$ by letting $\sigma(\omega)=\omega^k$ for some $k$
relatively prime to $\ell$ and stipulating that $\sigma(a)=a$ for all
$a\in F$. Then $\sigma$ is an automorphism of $K$ with $F$ as fixed
field and it generates the Galois group $\Gal(K/F)$.

The division algebra $D=(K/F,\sigma,z)$ is defined using a new
indeterminate $x$ as the $\ell$-dimensional vector space:
\[
D = K\oplus Kx\oplus \cdots \oplus Kx^{\ell-1},
\]
where the (noncommutative) multiplication for $D$ is defined by
$x^\ell = z$ and $xb = \sigma(b)x$ for all $b\in K$. Then $D$ is a
division algebra of dimension $\ell^2$ over $F$
\cite[Theorem 14.9]{Lam01}. The \emph{index} of $D$ is defined to be the square root of the dimension of $D$ over $F$. In our example, $D$ is of index $\ell$.

%The Galois automorphism $\sigma$ naturally extends from $D\rightarrow D$ by fixing $x$. In other words $\sigma(x)=x$. 
The elements of $D$ has matrix representation 
in $K^{\ell\times \ell}$ from its action on the basis 
$\mathcal{X}=\{1,x,\ldots,x^{\ell-1}\}$. I.e., for $a\in D$ and $x^j\in\mathcal{X}$, the $j^{th}$ row of the matrix representation is obtained by writing $x^{j} a$ in the $\mathcal{X}$-basis. 
%Its elements have matrix representations %in
%$K^{\ell \times \ell}$ (the regular matrix representation defined by
%multiplication from the left) given below:

For example, the matrix representation $M(x)$ of $x$ is:

\[
        M(x)[i,j] = \begin{dcases}
                        1 & \text{ if } j=i+1, i\le \ell-1 \\
                        z & \text{ if } i=\ell, j=1\\
                        0 & \text{ otherwise.}
                    \end{dcases}
\]

$$
M(x)=\begin{bmatrix}
    0       & 1 & 0 & \cdots & 0 \\
     0       & 0 & 1 &\cdots  &  0 \\
      \vdots & \vdots &\ddots &\ddots  &  \vdots \\
      0       & 0 &\cdots                  & 0 &1 \\
      z       & 0 & \cdots & 0 & 0
\end{bmatrix}.
$$

For each $b\in K$ its matrix representation $M(b)$ is:

\[
        M(b)[i,j] = \begin{dcases}
                        b & \text{ if } i=j=1 \\
                        \sigma^{i-1}(b) & \text{ if } i=j, i\ge 2\\
                        0 & \text{ otherwise.}
                    \end{dcases}
\]

\[M(b) = 
\begin{bmatrix}
b & 0 & 0 & 0 & 0 & 0  \\
0 & \sigma(b) & 0 & 0 & 0 & 0 \\
0 & 0 & \sigma^2(b) & 0 & 0 & 0 \\
0 & 0 & 0 & \ddots & 0 & 0 \\
0 & 0 & 0 & 0 & \sigma^{\ell-2}(b) & 0 \\
0 & 0 & 0 & 0 & 0 & \sigma^{\ell-1}(b)
\end{bmatrix}
\]
        
\begin{remark}
We note that $M(x)$ has a ``circulant'' matrix structure and $M(b)$ is
a diagonal matrix. For a vector $v\in K^\ell$, it is convenient to
write $\cir(v_1,v_2,\ldots,v_\ell)$ for the $\ell\times \ell$ matrix
with $(i,i+1)^{th}$ entry $v_i$ for $i\le \ell-1$, $(\ell,1)^{th}$
entry as $v_{\ell}$ and remaining entries zero. Thus, we have
$M(x)=\cir(1,1,\ldots,1,z)$.  Similarly, we write
$\diag(v_1,v_2,\ldots,v_\ell)$ for the diagonal matrix with entries
$v_i$.
\end{remark}

\begin{fact}
  The $F$-algebra generated by $M(x)$ and $M(b), b\in K$ is an
  isomorphic copy of the cyclic division algebra in the matrix algebra
  $\M_{\ell}(K)$.
\end{fact}

\begin{proposition}\label{circ-in-D}
  For all $b\in K$, $\cir(b,\sigma(b),\ldots,z\sigma^{\ell-1}(b)) = M(b)\cdot M(x)$.
\end{proposition}

Define $C_{i,j}= M(\omega^{j-1}) \cdot M(x^{i-1})$ for $1\leq i,j\leq \ell$. Observe that, $\B=\{C_{ij}, i,j \in [\ell]\}$ be a $F$-generating set for the division algebra $D$. The following proposition is a standard fact. 

%\begin{definition}\label{def:center}
%The center $\mathfrak{C}$ of the division algebra is the set of elements 
%$a\in D$ such that $a$ commutes with every %element in $D$. 
%\end{definition}
%It is easy to observe that $\mathfrak{C}$ is a field. Another standard fact is the following. 

\begin{proposition}\cite[Section 14(14.13)]{Lam01}\label{full-space}
Then $K$ linear span of $\B$ is the entire matrix algebra 
$\M_{\ell}(K)$. 
\end{proposition}

%\subsection*{Algebraic complexity}\label{alg-compl}
%\paragraph*{Algebraic branching program (ABP)}
%An algebraic branching program (ABP) is a layered directed acyclic graph with one in-degree-$0$ vertex called \emph{source}, and one out-degree $0$ vertex called \emph{sink}. Its vertex set is partitioned into layers $0, 1, \ldots, d$, with directed edges only between adjacent layers ($i$ to $i+1$). The source and the sink are in layers zero and $d$, respectively. Each edge is labeled by a linear form over $\F$ in variables $X = \{x_1,\ldots,x_n\}$. The polynomial computed by the ABP is the sum over all source-to-sink directed paths of the product of linear forms that label the edges of the path. The maximum number of nodes in any layer is called the width of the algebraic branching program. The size of the branching program is taken to be the total number of nodes.

%Equivalently, an ABP of width $w$ and $d$ many layers can be defined as an entry of a product of $d$ many linear matrices of size at most $w$. Therefore, the polynomial $f$ computed by an ABP is of form $(M_1\cdots M_d)_{i,j}$ for some $i,j\in [w]$.

\section{Noncommutative Rank Computation}\label{sect-idea}

In this section, we present the proof of Theorem~\ref{thm:main-theorem}. For the sake of the reader, let us first recall the definition of the inner rank, the blow-up rank of a linear matrix, and their equivalence from Section \ref{sec:overview}.

The noncommutative rank ($\ncrank$) or the inner rank of an $s\times s$ linear matrix $T$ over the noncommuting variables $x_1, x_2, \ldots, x_n$ is the minimum $r$ such that $T=P \cdot Q$ where $P$ is an $s\times r$ matrix and $Q$ is an $r\times s$ matrix with entries linear in $x_1, \ldots, x_n$~\cite{Coh95}.   

Let $T$ be an $s\times s$ matrix whose entries are linear forms over $\{x_1,x_2, \ldots,x_n\}$. We can write $T = A_0 + \sum_{i=1}^n A_i x_i$ where $A_0, A_1, \ldots, A_n$ are the coefficient matrices.   
Given such a matrix $T$ over the variables $x_1,\ldots, x_n$ and $d\in\mathbb{N}$, 
%=(g_{i,j})_{1\leq i,j\leq m}$ over 
%$\F\newbrak{x_1, \ldots, x_n}$ and $d\in \mathbb{N}$, 
let 
\[
T^{\{d\}}\ =\ \{T(\ubar{p}) \mid
 \ubar{p} \in {\M_d(\F)}^n\}.
\]  
Here $T(\ubar{p})=A_0\otimes I_d + \sum_{i=1}^n A_i\otimes p_i$. 
Define $\rank(T^{\{d\}})= \max_{\ubar{p}} \{\rank(T(\ubar{p}))\}.$  The regularity lemma \cite{IQS18, DM17} shows
that $\rank(T^{\{d\}})$ is always a multiple of $d$. 
In Section \ref{sec:construct-regular}, we discuss a constructive version of this lemma. 
 
%Here we provide another alternative proof of the regularity lemma.

%\begin{lemma}
%$\rank(T^{\{d\}})$ is always a multiple of $d$.
%\end{lemma}
%\begin{proof}
%We prove it by induction on the noncommutative rank of $T$.

%Consider a linear matrix $T$ of noncommutative rank 1. The base case trivially follows as any nonzero linear form of $T$ is nonzero for a scalar substitution. Therefore, for every $d$, $\rank(T^{\{d\}})$ is $d$ considering the diagonal matrices.

%Suppose the noncommutative rank of $T$ is $r$. It will have an $r\times r$ invertible sub-matrix $T'$. From the definition, $T'$ has an invertible sub-matrix $T''$ of size $r-1$. Up to some invertible linear transformations, we can write the following:
%\[
%T' = 
%\left(
%\begin{array}{c|c}
%T'' & v \\
%\hline
%u & w
%\end{array}
%\right).
%\]
%\end{proof}

\begin{definition}\label{defn:blow-up-rank}
The blow-up rank of the matrix $T$ is defined as 
\[
\ncrank^*(T)=\lim_{d\rightarrow \infty} \frac{\rank(T^{\{d\}})}{d}.
\]
\end{definition}
Using the regularity lemma and the weakly increasing property of the sequence $\left(\frac{\rank(T^{\{d\}})}{d}\right)_{d\geq 1}$, it is shown that the limit exists \cite[Chapter 4]{Makam18}. % using a very clean argument.
For any linear matrix $T$, it is known that $\ncrank(T)=\ncrank^{*}(T)$~\cite{IQS18}. Henceforth, we work with the blow-up rank of $T$ but continue to denote it by $\ncrank(T)$ for notational simplicity. The blow-up rank motivates us to define the following notion of a witness. 

%We first define the notion of a \emph{witness} of rank $r$. 
%The definition of the witness is motivated by the following fact~\cite{IQS18, DM17}.

%\begin{fact}
%Let $A_0, A_1, \ldots, A_n \in \F^{s\times s}$ and $T = A_0 + \sum_{i=1}^n A_ix_i$. The noncommutative rank of $T$ is $r$ if and only if for every sufficiently large $d$, the maximum rank obtained by evaluating $T$ over all $(p_1, \ldots, p_n) \in (\F^{d\times d})^n$ is $rd$.
%\end{fact}

\begin{definition}[Witness of rank $r$]
Let $A_0, A_1, \ldots, A_n \in \M_s(\F)$ and $T = A_0 + \sum_{i=1}^n A_ix_i$. 
%We say $(\alpha_1, \ldots, \alpha_n) \in \F^n$ is a witness of commutative rank $r$ of $T$ if $\rank(T(\alpha_1, \ldots, \alpha_n))\geq r$. 
We say that $\ubar{p}=(p_1, \ldots, p_n) \in \M_d(\F)^n$ for some $d$ is a witness of noncommutative rank $r$ of $T$, if $\rank(T(\ubar{p}))\geq rd$.
\end{definition}

\subsection{Constructive Regularity Lemma}\label{sec:construct-regular}

Suppose that for a linear matrix $T$, we already have a matrix tuple $\ubar{q}$ of dimension $d$, a witness of rank $r$ of $T$ such that $\rank(T(\ubar{q}))>rd$. Then the constructive regularity lemma offers
a simple and general procedure to get a $d\times d$ witness of rank $r+1$ for $T$ \cite{IQS18}. 
We state essentially the same proof as described in \cite{IQS18}. But for clarity and simplicity, we use the explicit cyclic division algebra construction described in Section~\ref{sec:cyclic}. Following Section~\ref{sec:cyclic}, the field $F=\Q(z)$ and $K=F(\omega)$.    
%The proof uses the construction of a central division algebra described in \cite[Lemma~5.7]{IQS18}. 

\begin{lemma}\label{lem:const-reg}\cite{IQS18}
For any $s\times s$ matrix $T = A_0 + \sum_{i=1}^n A_i x_i$, and a matrix tuple $\ubar{q}=(q_1,\ldots, q_n) \in \M_d(K)^n$ such that $\rank(T(\ubar{q})) > rd$, there exists a deterministic $\poly(n,s,d)$-time algorithm that returns another
matrix substitution $\ubar{q'}=(q'_1,\ldots, q'_n)\in \M_d(K)^n$ such that $\rank(T(\ubar{q'})) \geq (r+1)d$.
\end{lemma}

\begin{proof}

Let $D = (K/F,\sigma,z)$ be the cyclic division algebra described in Section~\ref{sec:cyclic}. Recall that $F = \Q(z)$ and $K = F(\omega)$ and $\B = \{C_{i,j} : i,j\in[d]\}$ is a $F$-generating set of $D$. 

\begin{enumerate}
\item Using Proposition \ref{full-space}, we can express $q_k = \sum_{i,j} \lambda_{i,j,k}C_{i,j}$ where $\lambda_{i,j,k} : 1\leq k\leq n$ are unknown variables which take values in $K$. Using linear algebra we determine the values $\lambda^{0}_{i,j,k} : 1\leq i,j\leq \ell, 1\leq k\leq n$ for the unknowns in $K$.\\

\item Now the goal is to compute a $d\times d$ tuple $\ubar{q'}=(q'_1,\ldots, q'_n)$ such that $q'_k= \sum_{i,j} \mu^{0}_{i,j,k}C_{i,j}$ where $\mu^{0}_{i,j,k}\in F$ and $\rank(T(\ubar{q'}))\geq (r+1)d$. 
We briefly describe the procedure outlined in~\cite{IQS18}. Write $\tilde{q}_1= \mu_{1,1,1} C_{1,1,1} + \sum_{(i,j)\neq (1,1)} \lambda^0_{i,j,1}C_{i,j}$ where $\mu_{1,1,1}$ is a variable. There will be a sub-matrix of size $>rd$ whose minor is non-zero, under the current substitution $(\tilde{q}_1, q_2, \ldots, q_n)$. Since the determinant of that sub-matrix is a univariate polynomial in $\mu_{1,1,1}$ and degree $\poly(r,d)$, we can easily fix the value of $\mu_{1,1,1}$ from $\Q$ such that the minor remains nonzero. Repeating the procedure, we can compute the desired tuple $\ubar{q'}$.      
Since $\ubar{q'}$ is a tuple over the division algebra, $\rank(T(\ubar{q'}))\geq (r+1)d$. 
\end{enumerate}
\end{proof}

\begin{remark}\label{rmk:rounding}
The last line of the above proof is easy to see. The matrix $T(\ubar{q'})$ can be viewed as a $s\times s$ block-matrix of $d$-dimensional blocks, and each such block is an element in $D$.   
%There will be a minor in $T(\ubar{q'})$ of rank $>rd$. 
Since Gaussian elimination is supported over division algebras, up to elementary row and column operations, we can transform $T(\ubar{q'})$ as:
\[
\left(
\begin{array}{c|c}
I & 0 \\
\hline
0 & 0
\end{array}
\right)
\]
where $I$ is an identity matrix which has at least $r+1$ blocks of identity matrices
$I_d$ on its diagonal. 
%\[
%I=\begin{bmatrix}
%I_{d}\\
%&I_{d}\\
%&&\ddots\\
%&&&&I_d
%\end{bmatrix}
%\]
%\end{remark}
%and the number of $I_d$ blocks is at least $r+1$. 
Hence $\rank(T(\ubar{q'}))\geq (r+1)d$. 
\end{remark}

\subsection{The Plan of the Algorithm}\label{sec:algo-plan}
Following \cite{IQS18}, we first give a simple template.

\begin{description}
\item[Algorithm Template]\

\textbf{Input:} $T = A_0 + \sum_{i=1}^n A_ix_i$ where $A_0, A_1, \ldots, A_n \in \M_s(\Q)$.

\textbf{Output:} The noncommutative rank of $T$.

\vspace{0.2cm}
%It is an incremental algorithm that 
The algorithm gradually 
constructs a witness at every stage. Suppose we already have a \emph{witness of rank $r$} for $T$. 
\vspace{0.1cm}
\begin{enumerate}
\item Is $r$ the maximum rank?

\item If yes, output $r$ to be the noncommutative rank of $T$.

\item Otherwise, find a witness of rank at least $r+1$ and go to Step~1.
\end{enumerate}
\end{description}
We now discuss each step in detail. 

\subsubsection{Rank Increment Step}\label{sec:rankincrement}

For an $s\times s$ linear matrix $T(\ubar{x}) = A_0 + \sum_{i=1}^n A_ix_i$ and $d\in\mathbb{N}$, define 
\[T_d(Z) = A_0\otimes I_d + \sum_{i=1}^n A_i\otimes Z_i\] 
where $Z_i = (z^{(i)}_{jk})_{1\leq j,k \leq d}$ is a $d\times d$ generic matrix with noncommutative indeterminates. In other words, $Z=(Z_1, Z_2, \ldots, Z_n)$ is the substitution used for the variables $x_1,x_2,\ldots,x_n$ in $T$. Now $T_d(Z)$ is a linear matrix of dimension $sd$ over the variables $\{z^{(i)}_{jk}\}_{1\leq j,k \leq d, 1\leq i\leq n}$. 
\begin{remark}\label{rmk:matrix-shift}
%Since we substitute the variables $x_1,\ldots,x_n$ by the $d\times d$ matrices $Z_1,\ldots,Z_n$, one can 
It is immediate to see that any $d\times d$ matrix shift $T_d(Z_1 + p_1, Z_2 + p_2,\ldots, Z_n + p_n)$ is indeed a scalar shift for the variables 
$\{z^{(i)}_{jk}\}_{1\leq j,k \leq d, 1\leq i\leq n}$ in the matrix $T_d$.
\end{remark}
%Intuitively, we would like to consider any $d\times d$ matrix shift of $T$ as a scalar substitution of $T_d$.

%For any integer $d'\in\mathbb{N}$, we define the bijective map $\iota_{d'}: \M_{d'}(K)^{nd^2} \to \M_{dd'}(K)^n$ as follows:
%\[
%\ubar{p}=(p^{(1)}_{11}, \ldots, p^{(1)}_{dd}, \ldots, p^{(n)}_{11}, \ldots, p^{(n)}_{dd})
%\mapsto \ubar{q}=(q_1, \ldots, q_n)
%\]
%where each $q_i = (p^{(i)}_{jk})_{1\leq j,k\leq d}$, i.e. we think of $q_i$ as $d\times d$ block matrix where the $(j,k)^{th}$ block is $p^{(i)}_{jk}$. 
%The inverse image is also well-defined.
%Notice that $T_{d}(\ubar{p})=A_0\otimes I_{dd'} + \sum_{i=1}^n A_i\otimes q_i = T(\ubar{q})$ where $\ubar{q}=\iota_{d'}(\ubar{p})$. 
%Notice that, for any matrix tuple $(q_1, \ldots, q_n)\in \M^n_{dd'}(\F)$, $T(\ubar{q}) = T_d(\iota^{-1}_{d'}(\ubar{q}))$.

\begin{lemma}\label{lem:scaling-blowup}
$\ncrank(T_{d}) = d \cdot \ncrank(T).$
\end{lemma}

\begin{proof}
Let $\ncrank(T) = r$. Therefore, for every sufficiently large $d''$, the maximum rank obtained by evaluating $T$ over all the  $d''\times d''$ matrix tuple is $rd''$. Consider $d'' = dd'$, a multiple of $d$ and let $\ubar{q}=(q_1, \ldots, q_n)\in\M_{d''}(K)^n$ be a matrix tuple such that $\rank(T(\ubar{q})) = rdd'$. Let 
\[
\ubar{p}=(p^{(1)}_{11}, \ldots, p^{(1)}_{dd}, \ldots, p^{(n)}_{11}, \ldots, p^{(n)}_{dd})
\]
be the matrix tuple in $\M_{d'}(K)^{nd^2}$ such that %$\iota_{d'}(\ubar{p}) = \ubar{q}$ where 
each $q_i = (p^{(i)}_{jk})_{1\leq j,k\leq d}$, i.e. we think of $q_i$ as $d\times d$ block matrix where the $(j,k)^{th}$ block is $p^{(i)}_{jk}$. Notice that $T_{d}(\ubar{p})=A_0\otimes I_{dd'} + \sum_{i=1}^n A_i\otimes q_i = T(\ubar{q})$. Notice that, the matrix $q_i$ is substituted for the variable $x_i$ in $T$. Therefore, $\rank(T(\ubar{q})) = \rank(T_d(\ubar{p}))$ and $\ncrank(T_d)\geq rd$.
%Suppose each $q_i = (p_{ijk})$, i.e. we think of $q_i$ as $d\times d$ block matrix where the $j,k$-th block is $p_{ijk}$. Consider evaluating $T_d$ by substituting $p_{ijk}$ for each $z_{ijk}$ variable. So we have a $d'\times d'$ substitution for the $z_{ijk}$ variables such that the rank is $rdd'$. 

%To show that $\ncrank(T_d)\leq rd$, assume to the contrary that $\ncrank(T_d)> rd$. Suppose, $\ncrank(T_d) = rd+k$ for some positive integer $k$. Therefore, for some $d'\times d'$ matrix tuple substitution $\ubar{p}\in\M_{d'}(K)^{nd^2}$, $\rank(T_d(\ubar{p})) = (rd+k)d' = rdd' + kd'$. From the definition, $\rank(T_d(\ubar{p})) = \rank(T(\iota_{d'}(\ubar{p}))) > rdd'$.
%Now define a $dd'\times dd'$ tuple $(q_1, \ldots, q_n)$ such that $q_i = (p_{ijk})$, i.e. we think of $q_i$ as $d\times d$ block matrix where the $j,k$-th block is $p_{ijk}$. Therefore, $\rank(T(\ubar{q})) = rdd' + kd'$. 
%However, $\ncrank(T) = r$ and for any $(q_1, \ldots, q_n)\in \M_{dd'}(K)^n$, $\rank(T(\ubar{q}))\leq rdd'$, hence a contradiction. Therefore, $\ncrank(T_d)\leq rd$.
For the other direction, as $\ncrank(T) = r$, we can write $T = P\cdot Q$ where $P, Q$ are $s\times r$ and $r\times s$ matrices respectively with linear entries. We can now define an $sd\times rd$ matrix $P'(Z)$ by substituting each $x_i$ by $Z_i$ in the matrix $P(\ubar{x})$. Similarly, we can define a $rd\times sd$ matrix $Q'(Z)$ from $Q(\ubar{x})$. Notice that, $T_d = P'\cdot Q'$. Therefore, $\ncrank(T_d)\leq rd$. Hence, the lemma follows. 
\end{proof}

Suppose, we have already computed a witness of noncommutative rank $r$ of $T$, namely $\ubar{p}=(p_1, \ldots, p_n) \in \M_d(K)^n$ (by construction, we will ensure that $d\leq r+1$). %If $\ncrank(T) > r$, we now construct a witness of rank $r+1$.
We now check whether $\ncrank(T) > r$ or not.
%\begin{observation}
%Evaluating $T$ at $(p_1, \ldots, p_n)$ is equivalent to evaluating $T^{\{d\}}$ by substituting each $z_{i,j,k}$ by $p_{i,j,k}$.    
%\end{observation}
\[
\text{Observe that, }T_d(Z_1 + p_1, \ldots, Z_n + p_n) =
U\left(
\begin{array}{c|c}
I_{rd} - L & A \\
\hline
B & C
\end{array}
\right)V
\]
for invertible transformations $U, V$ in $\M_{rd}(K)$. In fact, using further invertible transformations $U', V'$ over the free skew field $K\newbrak{Z}$ we can write 
\[
T_d(Z_1 + p_1, \ldots, Z_n + p_n) =
U U'\left(
\begin{array}{c|c}
I_{rd} - L & 0 \\
\hline
0 & C - B(I_{rd} - L)^{-1}A
\end{array}
\right) V'V. 
\]

\[
\text{Here,~}
U'= \left(
\begin{array}{c|c}
I_{rd}  & 0 \\
\hline
B (I_{rd} - L)^{-1} & I_{(s-r)d}
\end{array}
\right)
,
\quad\quad 
V'= \left(
\begin{array}{c|c}
I_{rd}  & (I_{rd}-L)^{-1}A \\
\hline
0 & I_{(s-r)d}
\end{array}
\right).
\]

Let $\widetilde{T_d}=C - B(I_{rd} - L)^{-1}A$.  
Notice that the $(i,j)^{th}$ entry of $\widetilde{T_d}$ is given by $\widetilde{(T_d)}_{ij} = C_{ij} - B_{i}(I_{rd} - L)^{-1}A_{j}$ where $B_i$ is the $i^{th}$ row vector of $B$ and $A_j$ is the $j^{th}$ column vector of $A$. 

\begin{lemma}\label{lemma:reduce-to-PIT}
$\ncrank(T) > r$ if and only if $\widetilde{(T_d)}_{ij}\neq 0$ for some choice of $i,j$.
\end{lemma}

\begin{proof}
 %Multiplying further by invertible matrices, we can write,
%\[
%T_d({Z}+\ubar{p}) =
%U\left(
%\begin{array}{c|c}
%I_{rd} - L & 0 \\
%\hline
%0 & C - B(I_{rd} - L)^{-1}A
%\end{array}
%\right) V. 
%\]
Let $\ncrank(T) > r$. Then by Lemma \ref{lem:scaling-blowup}, $\ncrank(T_d)>rd$. The noncommutative rank of a linear matrix is invariant under a scalar shift \footnote{Consider a linear matrix $L$ that achieves the maximum rank for a matrix substitution $\ubar{q}$ of some dimension $d$. Then, for any scalar shift $(\alpha_1,\ldots,\alpha_n)$, the shifted linear matrix $L(\ubar{x}+\ubar{\alpha})$ achieves the same rank on the matrix tuple $\ubar{q}-\ubar{\alpha}\otimes I_d$.}, 
hence $\ncrank(T_d(Z_1+p_1, \ldots, Z_n + p_n)) = \ncrank (T_d) > rd$. However, if $C - B(I_{rd} - L)^{-1}A$ is a zero matrix, this is impossible. 

Conversely if $\widetilde{(T_d)}_{ij} = C_{ij} - B_{i}(I_{rd} - L)^{-1}A_{j}$ is nonzero for some indices $i,j$, we can find matrix substitutions $\tilde{p}^{(k)}_{\ell_1\ell_2}$ of dimension $d'$ for the variables $\{{z}^{(k)}_{\ell_1\ell_2}\}_{1\leq \ell_1, \ell_2\leq d, 1\leq k\leq n}$, such that the rank of $T_d(Z_1+p_1, \ldots, Z_n+p_n)$ on that substitution is more than $rdd'$. Therefore, $\ncrank(T_d(Z_1+p_1, \ldots, Z_n+p_n)) > rd$.
%More precisely, let $\tilde{\ubar{p}}=\left(\tilde{p}^{\{k\}}_{\ell_1,\ell_2}\right)_{1\leq \ell_1, \ell_2\leq d, 1\leq k\leq n}$. 
%$\ubar{q}$ from sufficiently large dimension $dd'$ such that 
%Let $\ubar{q}=\iota_{d'}(\tilde{\ubar{p}})$ and $\ubar{p}\otimes I_{d'}=(p_1\otimes I_{d'}, \ldots, p_n\otimes I_{d'})$, then the rank of $T_d(\ubar{q}+\ubar{p}\otimes I_{d'})$ is more than $rdd'$. 
%This is same as saying the rank of $T_d(Z)$ is more than $rdd'$ when the variables $z^{(k)}_{\ell_1,\ell_2}$ is substituted by $\tilde{p}^{k}_{\ell_1,\ell_2} + p_k(\ell_1,\ell_2) \otimes I_{d'}$ for $1\leq k\leq n, 1\leq \ell_1, \ell_2\leq d$.
Hence $\ncrank(T_d)>rd$. 
By Lemma \ref{lem:scaling-blowup}, we get that $\ncrank(T)>r$. 
\end{proof}

The next lemma says that the infinite series $\widetilde{(T_d)}_{ij}\neq 0$ is equivalent in saying that a suitably truncated polynomial $\widetilde{P}_{ij}$ obtained from $\widetilde{(T_d)}_{ij}$ is nonzero. The proof of the lemma is fairly standard~\cite[Corollary 8.3, Page 145]{Eilenberg74}. However, we present a self-contained proof of it.

\begin{lemma}\label{lem:trucated}
$\widetilde{(T_d)}_{ij}\neq 0$ if and only if $\widetilde{P}_{ij}=C_{ij} - B_i\left(\sum_{k\leq rd-1} L^k\right)A_j\neq 0$.   
%the following is true:
%\[
%C_{ij} \neq 0,
%B_i A_j \neq 0,
%B_i L A_j \neq 0,
%\ldots,
%B_i L^{r - 1} A_j \neq 0.
%\]
\end{lemma}

\begin{proof}
We first notice that, $\widetilde{(T_d)}_{ij}$ is zero in 
the free skew field over the $Z$-variables 
%$\F\newbrak{Z}$ 
if and only if the formal power series  $C_{ij} - B_i\left(\sum_{k\geq 0} L^k\right)A_j$ is zero.
Therefore, if $\widetilde{(T_d)}_{ij}=0$, then obviously $\widetilde{P}_{ij}=0$ as the power series is zero.

Now suppose $\widetilde{P}_{ij}=0$. 
Note the terms in $C_{ij}$ are linear forms and the degree of any term in $B_i\left(\sum_{k\geq 0} L^k\right)A_j$ is at least two. So $C_{ij}$ must be zero. For simplicity, identify the $Z$-variables with $z_1, z_2, \ldots, z_N$ where $N=nd^2$. Write the row and column vectors $B_i$ and $A_j$ as $B_i=\sum_{\ell} B_{i,\ell} z_{\ell}, A_j=\sum_{\ell} A_{j,\ell} z_{\ell}$. Similarly, write 
$L=\sum_{\ell} L_{\ell} z_{\ell}$. Let, if possible, $B_i L^{rd} A_j$ contributes a nonzero monomial (word) $w=z_{i_1}z_{i_2}\ldots z_{i_{rd+2}}$. Clearly the coefficient of $w$ is $B_{i,i_1}L_{i_2}\ldots L_{i_{rd+1}}A_{j,i_{rd+2}}$. Look at the vectors  $v_1=B_{i,i_1}$, $v_2=B_{i,i_1} L_{i_2}$, \ldots, $v_{rd+1}=B_{i,i_1} L_{i_2}\ldots L_{i_{rd+1}}$ corresponding to the prefix words $w_1=z_{i_1}, w_2=z_{i_1}z_{i_2}, \ldots, w_{rd+1}=z_{i_1}\ldots z_{i_{rd+1}}$. They can not be all linearly independent since they are $rd$-dimensional vectors. Hence there exists scalars $\lambda_1,\ldots, \lambda_{rd+1}$ such that $\lambda_1 v_1 + \ldots \lambda_{rd+1} v_{rd+1}=0$. However, $v_{rd+1} A_{j,i_{rd+2}}\neq 0$ by the assumption. Hence there exists at least one vector $v_{\ell} : 1\leq \ell \leq rd$ such that $v_{\ell} A_{j,i_{rd+2}}\neq 0$. This means that the coefficient of the word $w_{\ell}z_{i_{rd+2}}$ of length at most $rd+1$ is nonzero in $\widetilde{P}_{ij}$, which is not possible.

We now repeat the same argument to show the infinite series $B_i \left(\sum_{k\geq 0} L^k\right) A_j=0$.
\end{proof}

Next, we apply Corollary \ref{cor:rs-ams-adapted} and Corollary \ref{cor:abp-witness} to output a matrix tuple efficiently on which $\widetilde{(T_d)}_{ij}$ evaluates to nonzero and $I_{rd}-L$ evaluates to a full rank matrix. 

\begin{lemma}\label{lem:nonzero-output}
There is a deterministic $\poly(n,r,d)$-time algorithm that can output a matrix tuple $\ubar{q}$ of dimension at most $d'=2rd$ for the ${Z}$ variables such that $I_{rdd'}-L(\ubar{q})$ is invertible and $\widetilde{(T_d)}_{ij}(\ubar{q})\neq 0$. 
%for some choice of $i,j$. Moreover, if $\widetilde{(T_d)}_{i,j}\neq 0$ for some $i,j$, it outputs a witness of its non-zeroness. 
\end{lemma}

\begin{proof}
 Notice that $\widetilde{P}_{ij}$ is an ABP of size $\poly(r,d)$ and the number of layers is at most $rd+1$. Applying Corollary \ref{cor:abp-witness}, we get a matrix tuple of dimension at most $rd+2$ such that $\widetilde{P}_{ij}$ evaluates on it to nonzero. By simple padding, we can get a matrix tuple $\ubar{q'}$ of dimension $d'=2rd$ such that $\widetilde{P}_{ij}(\ubar{q'})\neq 0$. Since $\ubar{q'}$ is a substitution for the $Z$ variables 
 $\{z^{(k)}_{\ell_1\ell_2}\}$ where $1\leq k\leq n, 1\leq \ell_1, \ell_2\leq d$, we write $\ubar{q'}=(q'^{(1)}_{11}, \ldots, q'^{(1)}_{dd}, \ldots, q'^{(n)}_{11}, \ldots, q'^{(n)}_{dd})$ for more clarity. Here each $q'^{(k)}_{\ell_1\ell_2}$ is a $d'$ dimensional matrix.  
 
 Consider a commutative variable $t$ and the scaled matrix tuple $t\ubar{q'}$. It is easy to see that the infinite series $C_{ij} - B_i (I_{rdd'}-L(t\ubar{q'}))^{-1} A_j$ is nonzero since the $k^{th}$ homogeneous part $t^k B_i L^k(\ubar{q'}) A_j$ will not mix with other homogeneous components.   
 
 However this also has a rational representation $\widetilde{(T_d)}_{ij}(t\ubar{q'})=\gamma_1(t)/\gamma_2(t)$ where $t$-degrees of the polynomials $\gamma_1(t), \gamma_2(t)$ are bounded by $rdd'$. Moreover 
 $I_{rdd'}-L(t\ubar{q'})$ is an invertible matrix and the degree of $\det(I_{rdd'}-L(t\ubar{q'}))$ is bounded by $rdd'$ over the variable $t$. Simply by varying the variable $t$ over a suitable large set $\Gamma$ of size $O(rd)$, we can fix a value for $t=t_0$ such that $\widetilde{(T_d)}_{ij}(t_0\ubar{q'})\neq 0$ and $I_{rdd'}-L(t_0\ubar{q'})$ is of rank $rdd'$. Define $\ubar{q}=t_0\ubar{q'}$.  
 %The proof follows by using Schutzenberger's theorem and the algorithm of Raz and Shpilka. RANDOM paper
\end{proof}

Following is an immediate corollary. 
\begin{corollary}\label{cor:immediate}
Suppose Lemma~\ref{lem:nonzero-output} outputs a matrix tuple $\ubar{q}$. We can compute another matrix tuple $\ubar{p'}$ of dimension $dd'$ which is a witness of $\ncrank(T)>r$. 
\end{corollary}
\begin{proof}
Define the matrix tuple 
$\ubar{q''}=(q''^{(1)}_{11}, \ldots, q''^{(1)}_{dd}, \ldots, q''^{(n)}_{11}, \ldots, q''^{(n)}_{dd})$ where $q''^{(k)}_{\ell_1\ell_2}=q^{(k)}_{\ell_1\ell_2} + p^{(k)}_{\ell_1\ell_2}\otimes I_{d'}$ is a $d'$ dimensional matrix tuple for $1\leq k\leq n, 1\leq \ell_1,\ell_2\leq d$.

Lemma \ref{lem:nonzero-output} shows that the rank of $T_d$ evaluated on the matrix tuple $\ubar{q''}$ is more than $rdd'$. This is same as saying that $T_d(Z)$ is of rank more than $rdd'$ when the variable $z^{k}_{\ell_1,\ell_2} : 1\leq k\leq n, 1\leq \ell_1, \ell_2\leq d$ is substituted by $q''^{(k)}_{\ell_1\ell_2}$. 
%\[
%\ubar{q'}+\ubar{p}\otimes I_{2r}=(q'^{(1)}_{11} + (p_1)_{11}\otimes I_{2r}, \ldots, q'^{(1)}_{dd} + (p_1)_{dd}\otimes I_{2r}, \ldots, q'^{(n)}_{11} + (p_n)_{11}\otimes I_{2r}, \ldots, 
%q'^{(n)}_{dd} + (p_n)_{dd} + \otimes I_{2r})
%\] 
%is more than $rdd'$.  
%(\ubar{q}+I_{2r}\otimes \ubar{p})>rdd'$. 
Hence $\ncrank(T_d)>rd$. By Lemma \ref{lem:scaling-blowup}, we know that $\ncrank(T)>r$.
Moreover, we obtain a matrix tuple $\ubar{p'}=({p'_1}, {p'_2}, \ldots, {p'_n})$ which is a witness of $\ncrank(T)>r$, where ${p'_k}=\left(q''^{(k)}_{\ell_1\ell_2}\right)_{1\leq \ell_1,\ell_2\leq d} : 1\leq k\leq n$. Notice that $\ubar{p'}$ is the substitution for the $\ubar{x}$ variables. 
%Moreover, the matrix tuple $\ubar{p'}=({p'_1}, {p'_2}, \ldots, {p'_n})$ for the $\ubar{x}$ variables, where ${p'_k}=\left(q''^{(k)}_{\ell_1\ell_2}\right)_{1\leq \ell_1,\ell_2\leq d}$ ($1\leq k\leq n$), a witness of $\ncrank(T)>r$. 
\end{proof}
%Equivalently, the matrices $(Z_1, Z_2, \ldots, Z_n)$ will be substituted by the $n$-tuple $\tilde{\ubar{q}}$ where 

%\textcolor{red}{Let $\iota_{d'}(\ubar{q'})=\tilde{\ubar{q}}$. 
%blow up each $z_{i,j,k} : 1\leq i\leq n, 1\leq j,k\leq d$ 
%variable by the matrix $q_{i,j,k}$ to construct a matrix tuple $\tilde{\ubar{{q}}}$ of dimension $dd'$. 
%It is easy to see that, $\ubar{q^{''}}=\tilde{\ubar{{q}}} + \ubar{p}\otimes I_{d'}$ is a substitution for the $\ubar{x}$ variables to witness that $\ncrank(T)>r$. }
%We can now construct $(p'_1, \ldots, p'_n)\in (\F^{d'\times d'})^n$ such that $\rank(T(\ubar{p'})) > rd'$. 

\subsubsection{Rounding and Blow-up Control}\label{sec:roundblowup}

Next, we apply Lemma \ref{lem:const-reg} which gives a rounding procedure to get a matrix tuple of dimension $d_1=dd'$ to witness that $\ncrank(T)=r'$ where $r' \geq r+1$. Call that new matrix tuple as $\ubar{p''}$. 

However, we can not afford to have such a dimension blow-up for the witness matrix tuple in every step of the iteration as it incurs an exponential blow-up in the dimension of the final witness. To control that, we use a simple trick from \cite{IQS18} which we describe for the sake of completeness.

\begin{lemma}\label{lem:blow-up-control}
Consider an $s\times s$ linear matrix $T$ and a matrix tuple $\ubar{p''}$ in $\M_{d_1}(K)^n$ such that $\ubar{p''}$ is a witness of rank $r'$ of $T$. We can efficiently compute another matrix tuple $\widehat{\ubar{p}}$ of dimension at most $r'+1$ such that $\widehat{\ubar{p}}$ is also a witness of rank $r'$ of $T$. 
\end{lemma}

\begin{proof}
%Let $d_1=dd'$. 
Consider a sub-matrix $A$ in $T(\ubar{p''})$ such that $\rank(A)$ is at least $r'd_1$. From each matrix in the tuple $\ubar{p''}$, remove the last row and the column to get another tuple $\widetilde{{\ubar{p}}}$. We claim that the corresponding sub-matrix $A'$ in $T(\widetilde{{\ubar{p}}})$ is of rank $>(r'-1)(d_1-1)$ as long as $d_1>r'+1$. Otherwise, $\rank(A)\leq \rank(A') + 2r'\leq (r'-1)(d_1-1)+2r' = r'd_1 - d_1 + r'+1 < r'd_1$. 
%Applying the procedure repeatedly, we can control the blow-up in the dimension. 
Now we can use the constructive regularity lemma (Lemma~\ref{lem:const-reg}) on the tuple $\widetilde{{\ubar{p}}}$ to obtain another witness of dimension $d_1-1$ which is a witness of rank $r'$ of $T$. Applying the procedure repeatedly, we can control the blow-up in the dimension within $r'+1$ and get the witness tuple $\widehat{\ubar{p}}$. 
\end{proof}

\subsection{The Final Algorithm}\label{sec:final-algorithm}
We now summarize our overall strategy. 

\begin{description}
\item[Algorithm]\

\textbf{Input:} $T = A_0 + \sum_{i=1}^n A_ix_i$ where $A_0, A_1, \ldots, A_n \in \M_s(\Q)$.

\textbf{Output:} The noncommutative rank of $T$.\\

%It is an incremental algorithm that constructs a witness at every stage. 
The algorithm gradually increases the rank and finds a witness for it. Suppose at any intermediate stage, we already have a matrix tuple $\ubar{p}$ in $\M_d(K)^n$, a \emph{witness of rank $r$} of $T$. 

\begin{enumerate}
\item (Is $r$ the maximum rank?) Use Theorem \ref{razshpilka} to check whether the polynomial $\widetilde{P}_{ij}\neq 0$ (as defined in Lemma \ref{lem:trucated}) for some choice of $i,j$.
\item If "NO", output $r$ to be the noncommutative rank of $T$.
\item (Otherwise, construct a witness of rank $r+1$ and repeat Step~1) We implement the following steps to construct a rank $(r+1)$-witness:
\begin{enumerate}
\item~[Rank increment step] Apply Corollary \ref{cor:immediate}
%Section \ref{sec:rankincrement} 
to find a $d_1\times d_1$ matrix substitution $\ubar{p'}=(p'_1,\ldots, p'_n)$ such that $\rank(T(\ubar{p'})) > rd_1$ where $d_1=2rd^2$. 
%(from Step~1).

\item~[Rounding using the regularity lemma] Apply Lemma \ref{lem:const-reg} to find another $d_1\times d_1$ matrix substitution $(p''_1, \ldots, p''_n)$ such that the rank of $T$ evaluated at $(p''_1, \ldots, p''_n)$ is $r'd_1$ where $r'\geq r+1$. %[Using the constructive version of regularity lemma as discussed in Section \ref{sec:construct-regular} and in Section \ref{sec:roundblowup}].

\item ~[Reducing the witness size] Apply Lemma \ref{lem:blow-up-control} to find a matrix substitution $\widehat{\ubar{p}}=(\widehat{p}_1, \ldots, \widehat{p}_n)$ of dimension $d' \leq r'+1$, such that the rank of $T$ evaluated at $\widehat{\ubar{p}}$ is $\geq r'd'$. 
%[Using the blow-up control operation described in Section \ref{sec:roundblowup}].
\end{enumerate}
\end{enumerate}
\end{description}

\begin{description}
\item[Analysis] 
\end{description}
Since the noncommutative rank of $T$ is at most $s$, the algorithm iterates at most $s$ steps. Lemma \ref{lem:trucated}, Theotem \ref{razshpilka}, and Lemma \ref{lem:nonzero-output} guarantee that Step 1 and Step 3(a) can be done in $\poly(n,r,d)$ steps. Step 3(b) and 3(c) require straightforward linear algebraic computations discussed in Section \ref{sec:roundblowup} which can be performed in $\poly(n,d,r)$ time. Since $d\leq s+1$ throughout the process, the run time is bounded by $\poly(n,s)$. Understanding the bit-complexity of the algorithm is very simple. Let the witness of rank $r$ has bit-complexity $b$. In the rank increment step the matrix constructed in Corollary \ref{cor:abp-witness} has only $0,1$ entries and the parameter $t_0$ is of $\poly(s,d)$. So the bit-complexity after step 3(a) can change to $b + \log(sd)$ at most. Step 3(b) is a simple linear algebraic step that can at most incur bit-complexity by an additive factor $\poly(s,d)$. Therefore, the overall bit-complexity of the algorithm is $\poly(s)$.    

\section{Further Remarks}\label{sec:conclusion}
It is well-known that testing whether a bipartite graph has a perfect matching can be reduced to $\nsing$ using Hall's theorem \cite{GGOW20}. The bipartite matching problem is in (black-box) quasi-NC via a hitting set construction \cite{FGT21}. In contrast, designing an efficient black-box algorithm (or even a parallel algorithm) for $\nsing$ is wide open \cite{GGOW20}. 
%understanding the black-box algorithm and the parallel complexity of $\nsing$ are wide open problems \cite{GGOW20}. 

Interestingly, the algorithm presented in this paper can be adapted to get a black-box solution for the following problem.% derived essentially from the main algorithm in Section \ref{sec:final-algorithm}.\\

\noindent\fbox{
    \parbox{\textwidth}{
        Given a $4$-tuple $\angle{s, r, d, n}\in \mathbb{N}^4$, construct efficiently a universal set $\mathcal{H}_{s,r,n,d}\subseteq\M_{krd}(\Q)^n$ (for some integer $k$) of sub-exponential size such that the following is true: \\
        
        Consider any $\Q$-linear matrix $T$ defined over ${x_1,x_2,\ldots,x_n}$ and a tuple $\ubar{p}\in\M_d(\Q)^n$ such that $\rank(T(\ubar{p}))\geq rd$. Then, $\ncrank(T) > r$ if and only if there exists a tuple $\ubar{q}\in\mathcal{H}_{s,r,n,d}$ such that $\rank(T(\ubar{q} + \ubar{p}\otimes I_{kr}))>rd$. %unless $rd$ is the maximum rank possible for $T(\ubar{p})$ when $\ubar{p}\in\M_d(\Q)^n$.
}}

\vspace{0.3cm}
The connection with the noncommutative ABP identity testing enables us to construct such a set of quasipolynomial-size. We need to use the quasipolynomial-size hitting set construction for noncommutative ABPs by Forbes and Shpilka \cite{FS13}. More precisely, in the proof of Lemma \ref{lem:nonzero-output} we need to use the hitting set of \cite{FS13} in place of Corollary \ref{cor:immediate}. The rest of the arguments remain exactly the same.  The size of the set 
$\mathcal{H}_{s,r,n,d}$ will be $(srnd)^{O(\log rd)}$ and the parameter $k$ will be $2rd$. 
This step is black-box in the sense that it does not use the explicit description of the coefficient matrices $A_0, A_1, A_2, \ldots, A_n$.    
%To the best of our understanding, we do not see 
It is unclear how to solve this problem using the second Wong sequence-based techniques \cite{IKQS15, IQS18}. 
%\newpage
\bibliographystyle{unsrt}
\bibliography{ref2}
\end{document}